\documentclass[twoside]{dis08}
\usepackage[latin1]{inputenc}
\usepackage[dvips]{graphicx,epsfig,color}
\usepackage{wrapfig,rotating}
\usepackage{amssymb,amsmath,array}

\begin{document}
\title{Measurement of $\Delta G/G$
from high transverse momentum hadron pairs
in COMPASS}

\author{Marcin Stolarski \\ on behalf of COMPASS collaboration
%
%
\vspace{.3cm}\\
%
CERN\\
CH-1211 Geneve 23, Switzerland
%
}

\maketitle

\begin{abstract}
The new COMPASS $\Delta G/G$ result obtained from
high transverse momentum hadron pairs 
in the $Q^{2}>1$ (GeV/$c$)$^{2}$ region is presented.
Comparing to the previous analysis in this region
the statistical error of $\Delta G/G$ is reduced by a factor 3
to 0.10.
A weighted method of the $\Delta G/G$
measurement based on neural network approach is used.
In addition, the formula for the $\Delta G/G$ extraction
used in the analysis has been updated.  The contributions
coming from the leading order and QCD Compton processes
are no longer neglected. Slides can be found in \cite{url}.

\end{abstract}

\section{Introduction}

COMPASS is an experiment at CERN
focusing on the spin structure of the nucleon and
hadron spectroscopy.
Between 2002 and 2004, a 160 GeV/c polarized
muon beam and a two cell polarized
$^6$LiD target were used 
for spin
studies. For the 2006 data taking major updates of the spectrometer were
made. 

Here we present results for $\Delta G/G$ extracted
from high transverse momentum hadrons pairs.
The selection of these events increases
the fraction of the {\em photo-gluon fusion} (PGF) in the sample, therefore
increases sensitivity to the gluon polarization.
So far data from 2002-2004 years were analyzed. 
The analysis of the 2006 data is in progress.
The $\Delta G/G$ value was extracted for both the $Q^{2}>1$ (GeV/c)$^{2}$ and the $Q^{2}<1$ 
(GeV/c)$^{2}$ regions. 
In this paper we present the new result for the $Q^{2}>1$ (GeV/c)$^{2}$
analysis. For  the low Q$^{2}$ data the obtained result
is  $\Delta G/G = 0.016 \pm 0.058 \pm 0.054$ $c.f.$ \cite{compass2}.
Similar analyzes where performed in SMC \cite{smc} and HERMES \cite{hermes}.
In addition, COMPASS measured $\Delta G/G$ for open charm events,
which are believed to be a clean source of PGF. 
The description of that analysis can be found in this proceedings, 
$c.f.$ contribution by Florent Robinet. 

\section{Data selection}

An incoming and a scattered muon as
well as an interaction vertex in the target are required
for each event.
The kinematic cuts $0.1<y<0.9$ and
$Q^{2}>1$ (GeV/c)$^{2}$ are applied.
The latter cut ensures that scale of the hard process is high enough so that pQCD can be used.
We require at least two charged hadrons in the interaction vertex
with transverse momentum above 0.7~GeV/c each, their
$x_{F}$ as well as $z$ have to be larger than 0.
In addition, is also required that the sum of the energy
fraction of the two hadrons is $z_{1}+z_{2}<0.95$
and that their invariant mass is $> 1.5$ GeV/c$^{2}$.
The total number of selected events  
is about 500k.

\section{Extraction of $\Delta G/G$} 

In the leading order QCD there are three sub-processes which contribute
to the total cross-section: leading order (LO), QCD Compton (QCDC or C),
and photo gluon-fusion (PGF). The cross-section asymmetry for 
high transverse momentum hadrons pair can be expressed as
 \begin{displaymath}
A_{LL}^{2h}(x_{Bj}) \approx \frac{\Delta G}{G}(x_{G}) \hat{a_{LL}^{PGF}} R_{PGF} + 
A_{1}^{LO}(x_{C})
\hat{a_{LL}^{C}} R_{C} + A_{1}^{LO}DR_{L};  \hspace{0.5cm}   A_{1}^{LO} \equiv \frac{\sum_{i} 
e_{i}^{2} \Delta q_{i}}{ \sum_{i} e_{i}^{2} q_{i}}
\end{displaymath}
where $a_{LL,i}$ are analyzing powers and $R_{i}$ are the fractions of 
sub-processes in the
sample. In a similar way the cross-section asymmetry $A_{1}^{d}$ 
can be decomposed.
Combining the two asymmetries, the following relation holds:
 \begin{displaymath}
   \frac{\Delta G}{G} (\bar{x_{G}}) =
 \frac{A_{LL}^{2h}(x_{Bj}) + A^{corr}}{\beta}    
 \end{displaymath}
 \begin{displaymath}
  \beta =  a_{LL}^{PGF} R_{PGF}  - a_{LL}^{PGF,incl}
R_{PGF}^{incl}
 \left(\frac{R_{L}}{R_{L}^{incl}} + \frac{R_{C}}{R_{L}^{incl}} \frac{a_{LL}^{C}}{D} 
\right)
 \end{displaymath}
 \begin{displaymath}
A^{corr} =  -A_{1} (x_{Bj})D \frac{R_L}{R_L^{incl}} -  A_{1}(x_{C})\beta_{1} + 
A_{1}(x'_{C}) 
\beta_{2}
 \end{displaymath}
 \begin{displaymath}
\beta_{1} =   \frac{1}{R_{L}^{incl}} \left( a_{LL}^{C} R_{C} - a_{LL}^{C,incl} R_{C}^{incl} 
\frac{R_{L}}{R_{L}^{incl}} \right) \; \; \;
\beta_{2} =   a_{LL}^{C,incl} \frac{R_{C}R_{C}^{incl}}{( R_{L}^{incl})^{2}} \frac{a_{LL}^{C}}{D}.
 \end{displaymath}

There are two points worth mentioning about the formulas.
Firstly, a $A_{corr}$ cannot be neglected for the $\Delta G/G$ extraction.
It was found that the average value of $x_{C}$ in the sample is about
0.14. For such a high $x$ value $A_{1}^{d}$ and therefore $A_{corr}$ are no longer zero.
Secondly, $\beta$ determines
the precision of the measured $\Delta G/G$. In the first approximation
it is proportional to a difference
between the PGF fractions for the high-$p_{T}$ and the inclusive samples.
If the difference between the two fractions is small
the statistical error of $\Delta G/G$
will be large. Neglecting the impact of PGF events from the inclusive 
sample would lead
to underestimation of the statistical error of $\Delta G/G$.

To increase the statistical precision of the gluon polarization
measurement a weighted method of asymmetry extraction was used. 
mThe weight is $fDP_{b} \beta$,
where $f$ is the dilution factor $D$ is the depolarization factor, $P_{b}$ 
is the beam polarization. The weight has to be known on the event
by event basis. This means that for every event we have to know:
$R_{PGF}, R_{C}, R_{L}, R_{PGF}^{incl}$, $R_{C}^{incl}, R_{L}^{incl},
 a_{LL}^{PGF}, a_{LL}^{C}$, $a_{LL}^{PGF,incl}, a_{LL}^{C,incl},
 x_{G}, x_{C},f, D, P_{b}$. Only the last three quantities are calculated
from data, all others have to be obtained from MC. 
A Neural Network (NN) trained on MC samples was used for the parametrization of 
these
quantities. As an input to the NN training for the inclusive sample, 
$x_{Bj}$ and $Q^{2}$ were selected while for the high$-p_{T}$ sample in 
addition transverse and longitudinal momenta of the two hadrons were used. 
This method of $\Delta G/G$ extraction  depends largely on MC.
Good data description with the MC as well as good NN parameterizations are 
the ``key points'' of this analysis.

\section{MC and NN parameterizations}

\begin{figure}[!h]
\begin{minipage}[h]{0.12\textwidth}
\hspace{0.3cm}
\end{minipage}
\begin{minipage}[h]{0.25\textwidth}
\centerline{\includegraphics[width=3.0cm]{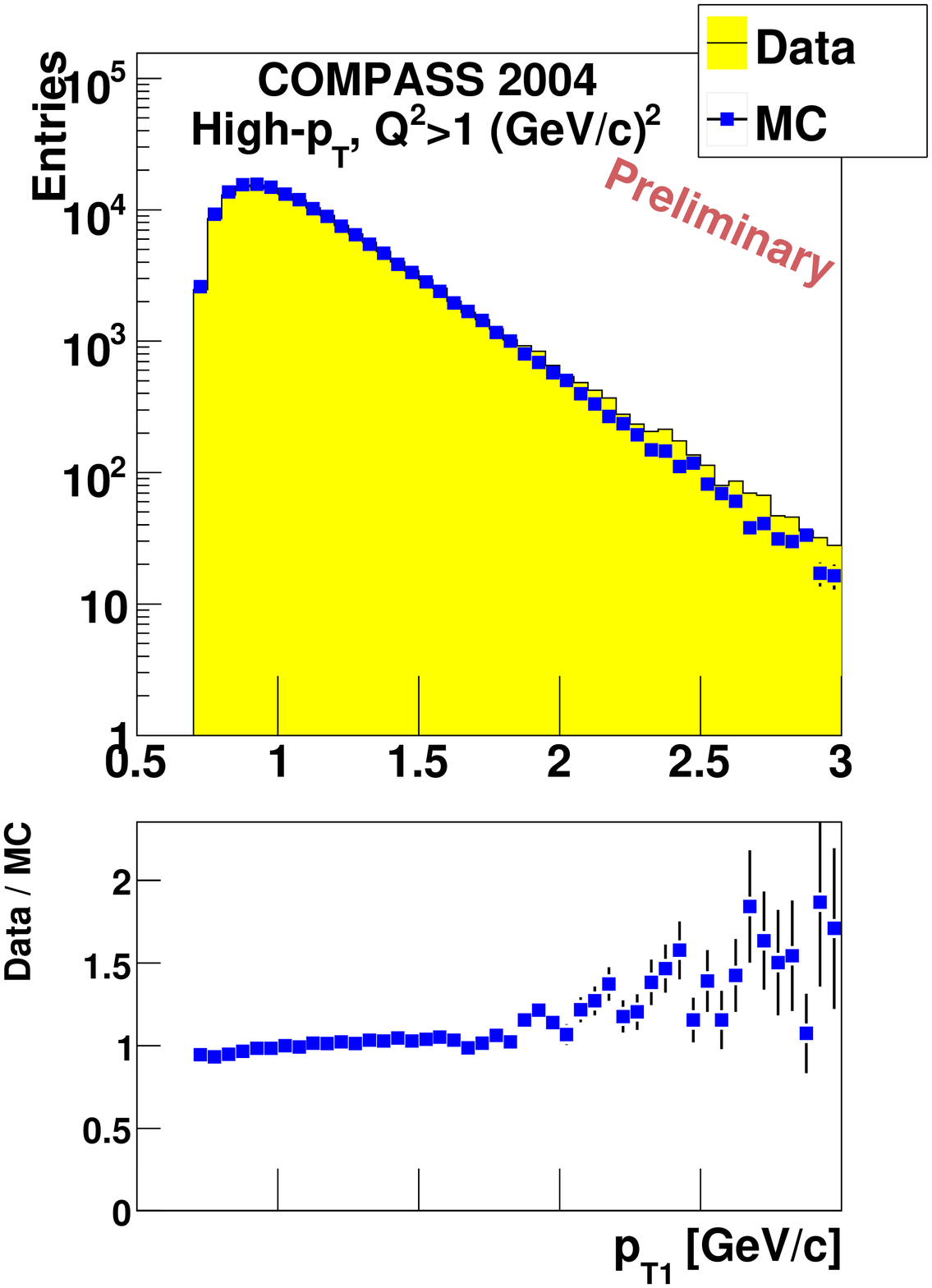}}
\end{minipage}
\begin{minipage}[h]{0.25\textwidth}
\centerline{\includegraphics[width=3.0cm]{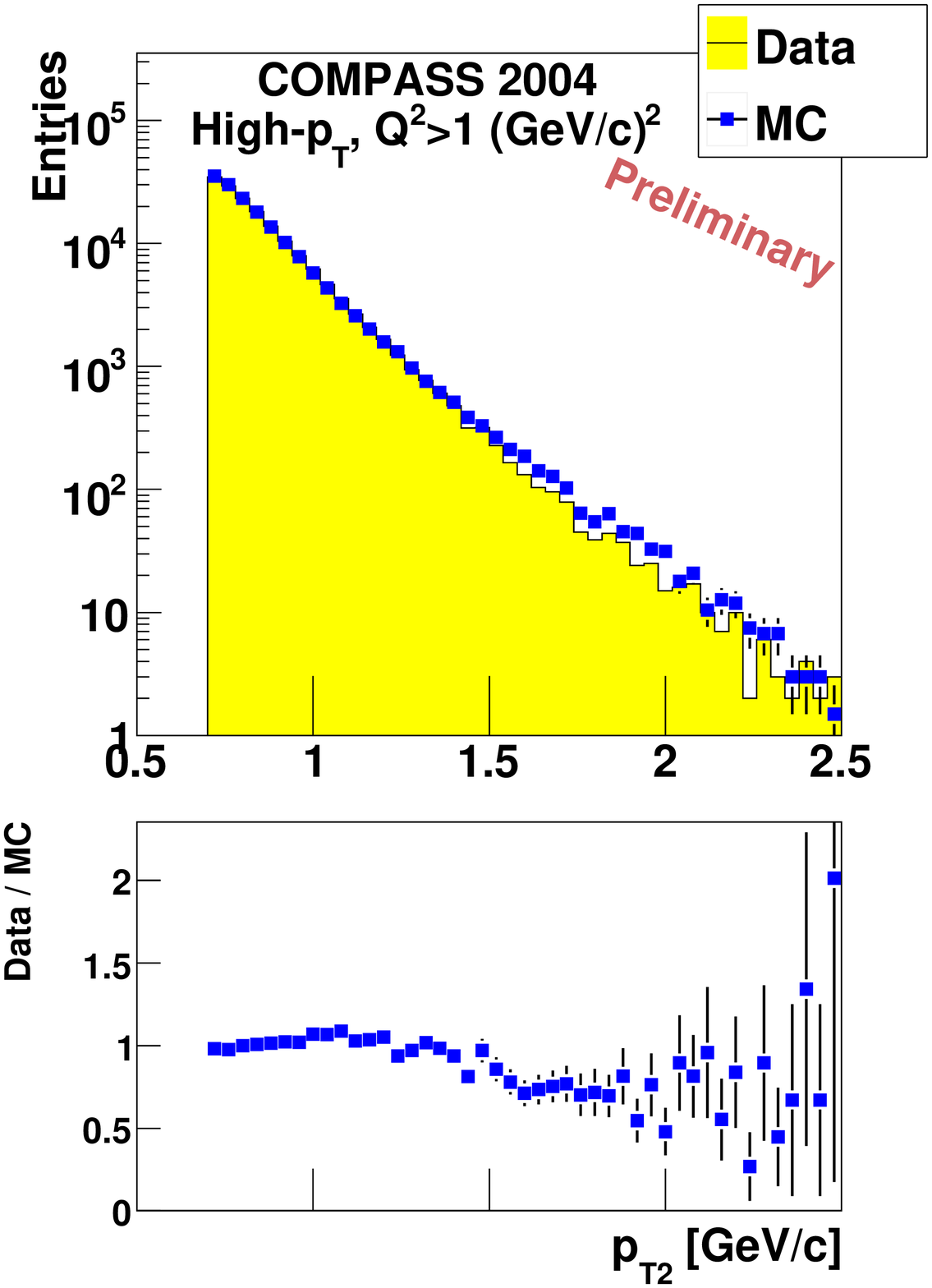}}
\end{minipage}
\begin{minipage}[h]{0.25\textwidth}
\centerline{\includegraphics[width=3.0cm]{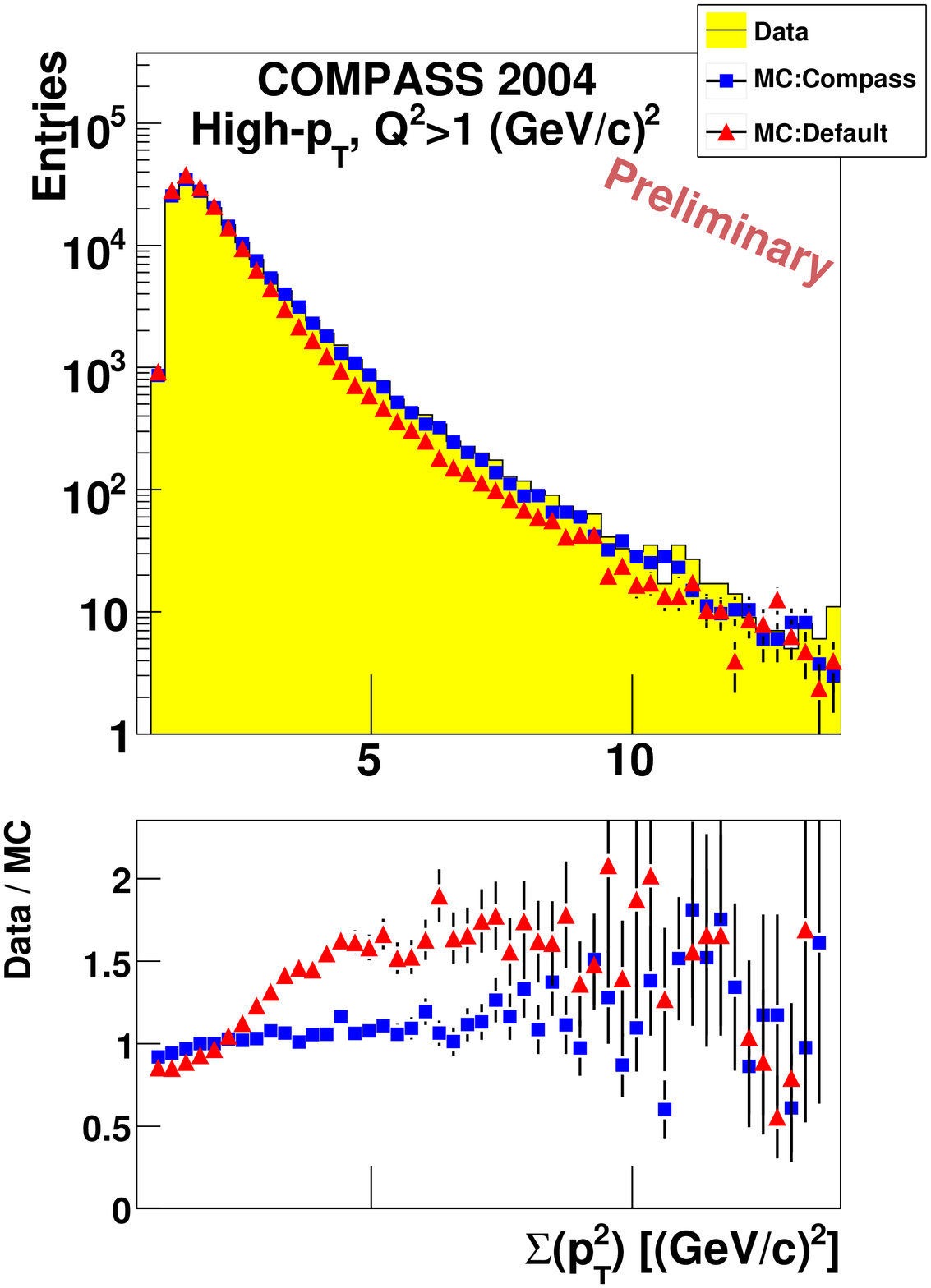}}
\end{minipage}
\caption{Data and Monte Carlo comparison for transverse momenta of 
the hadrons.}  \label{fig::mccompar}
\end{figure}

To extract $\Delta G/G$ two MC samples are needed:
an inclusive and a high-$p_{T}$.
The LEPTO generator and full simulation of the COMPASS
spectrometer were used. 
MRST04 LO was used for the  parton distribution functions given the very
good description of  $F_{2}$ in COMPASS
kinematics.
The gluon radiation in the initial and final state
were allowed in the generator (the so called parton shower ON). This 
simulates
part of NLO corrections. The parton shower OFF was used
to estimate systematics. To improve the data and MC agreement
the LEPTO parameters were tuned ($k_{T}$ and parameters of fragmentation).
Standard LEPTO tuning was used for systematic studies.
In Figure \ref{fig::mccompar} examples of data-MC comparison  are shown
for the transverse
momenta of the first and the second
hadron respectively. In the right-most plot the comparison between
standard and tuned LEPTO is shown for $\sum p_{T}^{2}$. We observe good agreement
between data and MC, also necessity for LEPTO tuning is clearly visible.
\vspace{-0.3cm}

\begin{figure}[!h]
\centerline{\includegraphics[width=9.5cm]{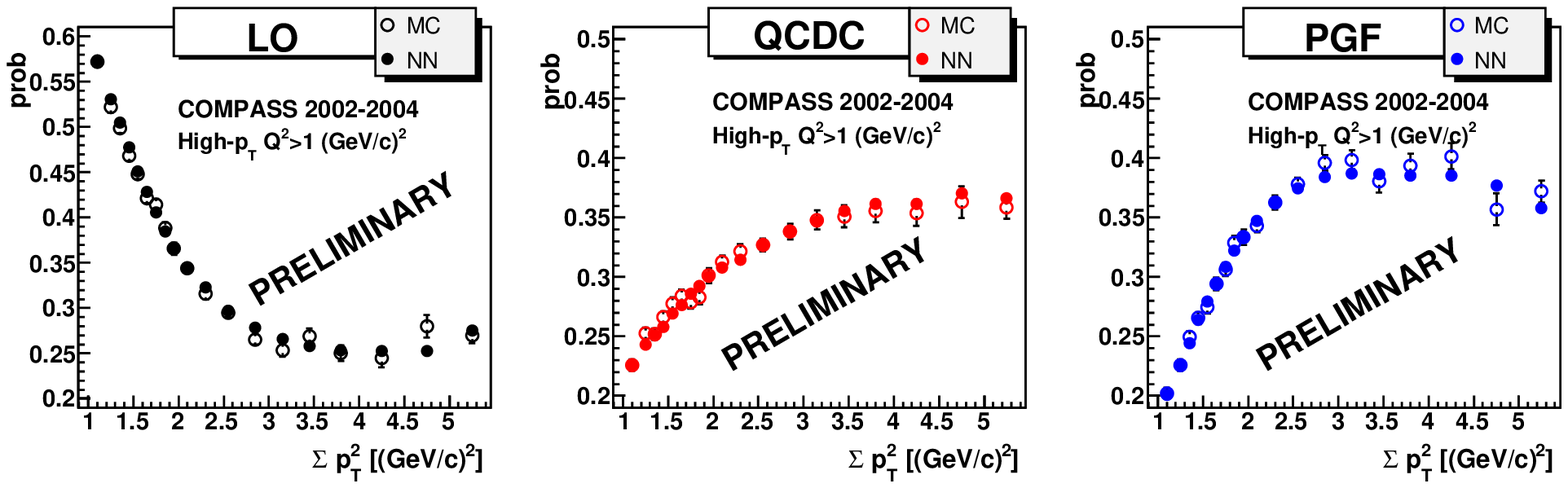}}
\centerline{\includegraphics[width=9.5cm]{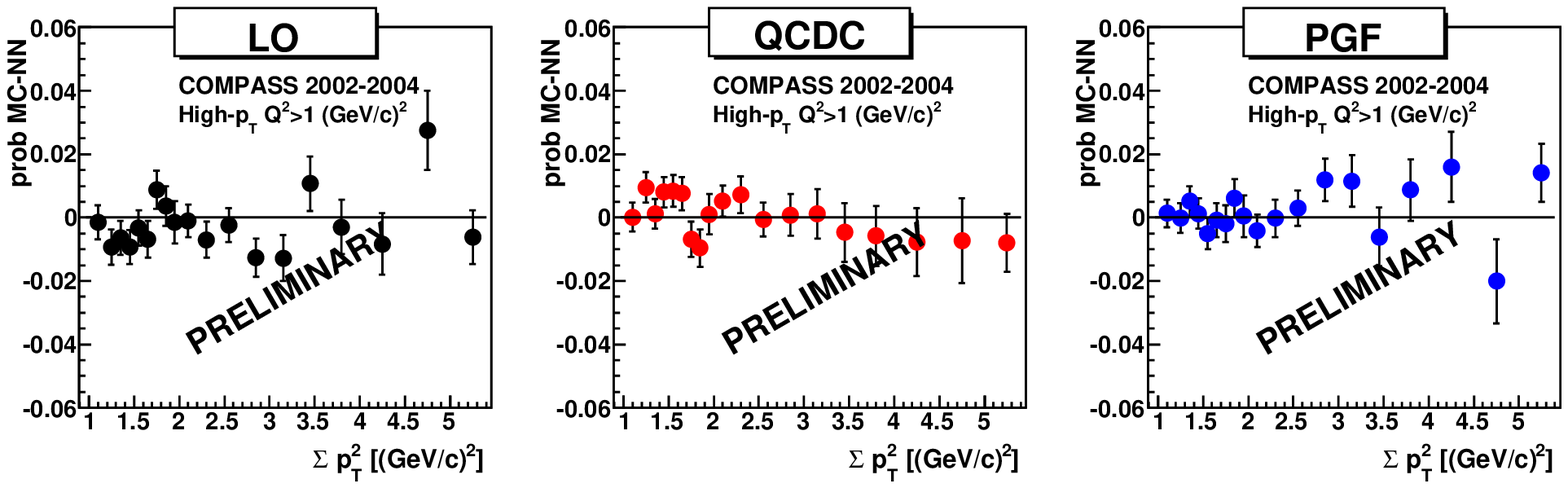}}
\caption{ Comparison of $R_{i}$ from MC and NN parametrization.} \label{fig::nn1}
\end{figure}

The NN parameterizations of the variables used in the weight
were compared with original MC samples and good agreement
was found. As an example we show
in Figure \ref{fig::nn1} the comparison between the fractions of
the processes as a function of $\sum p_{T}^{2}.$

\section{Systematic studies}

Various systematic studies were performed to estimate
the systematic uncertainty of the extracted $\Delta G/G$.
Summary is given in Table \ref{tab::sys}. The main contribution
comes from the MC, other contributions like false asymmetries,
uncertainties of the beam/target polarization, dilution factor,
radiative correction, simplification of the formula for $\Delta G/G$ extraction,
$A_{1}^{d}$ parametrization and NN parametrization stability were found to 
be much smaller. 

\begin{wraptable}{l}{0.35\columnwidth}
\begin{center}
\begin{tabular}{|c|c|}
\hline
$\delta(\Delta G/G)_{NN}    $  &  0.006  \\ \hline
$\delta(\Delta G/G)_{MC}    $  &  0.040  \\ \hline
$\delta(\Delta G/G)_{f,P_{b},P_{t}} $& 0.006  \\ \hline
$\delta(\Delta G/G)_{false}  $ & 0.011  \\ \hline
$\delta(\Delta G/G)_{A1}     $ & 0.008  \\ \hline
$\delta(\Delta G/G)_{formula}$ & 0.013  \\ \hline \hline
TOTAL  &   0.045  \\ \hline
\end{tabular}
\caption{Contributions to the systematic error of $\Delta G/G$} \label{tab::sys}
\end{center}
\end{wraptable}
Since the MC error gives the largest contribution
to the systematic error its estimation will be described in
more details. For these studies we were using four MC samples:
tuned LEPTO with parton shower ON/OFF and standard
LEPTO tuning with parton shower ON/OFF. This way we 
estimated the impact of the MC tuning and partially the impact of NLO corrections. 
In addition, we made three extraction of $\Delta G/G$ with each of MC samples. 
In the first one
we were using samples as they were, 
in the second one using NN we selected events from regions of phase space
where data and MC agrees well 
in the last one variation
we re-weighted MC events so that ratio data over MC is equal to 1. 
This way we obtained 12 values of $\Delta G/G$. 
From their 
distribution we estimated the MC systematic error 
to be about 0.04.

\section{Results}
The preliminary result for $\Delta G/G$ for high transverse
momentum hadron pairs with $Q^{2}>1$ (GeV/c)$^{2}$  is $\Delta G/G = 0.08 \pm 0.10 \pm 0.05$.
The average $<x_{G}>$ for the data was about 0.08, and average scale of 
the 
process about 3 (GeV/c)$^{2}$. The result is consistent with zero. 
The statistical precision with respect to the previous measurement from COMPASS in the $Q^{2}>1$ 
(GeV/c)$^{2}$
range increased by factor 3. The presented result is also in good 
agreement
with the $\Delta G/G$ result obtain for low 
the $Q^{2}$ high-$p_{T}$ analysis.

\section{Summary and Outlook} 
The preliminary result for $\Delta G/G$ for high transverse
momentum hadron pairs with $Q^{2}>1$ (GeV/c)$^{2}$, was presented:
$\Delta G/G = 0.08 \pm 0.10 \pm 0.05$. The result is compatible
with zero as well as with other direct measurements of $\Delta G/G$
from COMPASS, HERMES and SMC. The presented data used 2002-2004
data. With additional 2006 data and future upgrade of the analysis method
we expect to at least double available statistics.



\begin{footnotesize}

\end{footnotesize}


\end{document}